\documentclass[12pt,fleqn,draft]{article}

\usepackage{a4wide}
\usepackage{cite}

\usepackage{amsmath,amstext}
\usepackage{amsfonts}
\usepackage{amsthm}
\usepackage{amscd}

\DeclareMathOperator{\fstr}{Str}
\DeclareMathOperator{\sn}{sn}
\DeclareMathOperator{\cn}{cn}
\DeclareMathOperator{\dn}{dn}
\DeclareMathOperator{\cd}{cd}
\DeclareMathOperator{\bsc}{sc}
\begin{document}
%%%%%%%%%%%%%%% TITLE %%%%%%%%%%%%%%%%%%%%%%%%%%%%%%%%%%%
\title{Fermionic $R$-Operator 
and Integrability of the One-Dimensional Hubbard Model}
         
\maketitle
\vspace{5mm}
%%%%%%%%%%%%%%%%%%%%%%%% AUTHOR(S) %%%%%%%%%%%%%%%%%%%%%%%%
\begin{center}
       Yukiko \textsc{Umeno}
        \footnote[2]{E-mail: 
                        \texttt{umeno@monet.phys.s.u-tokyo.ac.jp}
                   },
        Masahiro \textsc{Shiroishi}
        \footnote[8]{E-mail:
                        \texttt{siroisi@monet.phys.s.u-tokyo.ac.jp}
                   }   
        and
        Miki \textsc{Wadati}
%        \footnote[9]{E-mail:
%                       \texttt{wadati@monet.phys.s.u-tokyo.ac.jp}
%                  } 
\\
%%%%%%%%%%%%%%%%%%%%% ADDRESS %%%%%%%%%%%%%%%%%%%%%%%%%
         Department of Physics, Graduate School of Science,\\
         University of Tokyo,\\
         Hongo 7-3-1, Bunkyo-ku, Tokyo 113-0033, Japan
\end{center}
\date{}
%%%%%%%%%%%%%%%%%%%%%%%%%%%%%%%%%%%%%%%%%%%%%%%%%%%%%%%%%%%%%%%%%%%%%%%%%%%%%%
\vspace{5mm}

We propose a new type of the Yang-Baxter equation (YBE) and 
the decorated Yang-Baxter equation (DYBE). 
Those relations for the fermionic ${R}$-operator
were introduced recently as a tool to treat 
the integrability of the fermion models. 
Using the YBE and the DYBE for the ${XX}$ fermion model, 
we construct the fermionic ${R}$-operator 
for the one-dimensional (1D) Hubbard model. 
It gives another proof of the integrability of the 1D Hubbard model. 
Furthermore a new approach to the ${SO(4)}$ symmetry of the 1D Hubbard model
is discussed.
\vspace{5mm}

%%%%%%%%%%%%% KEYWORDS %%%%%%%%%%%%%%%%%%%%%%%%%%
\begin{center}
          KEYWORDS
\\
fermion ${XYZ}$ model, one-dimensional Hubbard model, fermionic ${R}$-operator, 
\\
Yang-Baxter equation, ${SO(4)}$ symmetry
 
\end{center}

%%%%%%%%%%%%%%%%%%%%%%%%%%%%%%%%%%%%%%%%%%%%%%
\section{Introduction}
\setcounter{equation}{0}
\renewcommand{\theequation}{1.\arabic{equation}}

There have been much interests in
strongly correlated electron systems.
Among the exactly solvable ones,
the one-dimensional (1D) Hubbard model,
\begin{align}
\mathcal{H} = - \sum_{j=1}^N \sum_{\sigma = \uparrow \downarrow}     
       (c_{j+1 \sigma}^{\dagger} c_{j \sigma} 
                  + c_{j \sigma}^{\dagger} c_{j+1 \sigma})  
       + U \sum_{j=1}^N   
    (n_{j \uparrow}-\frac{1}{2})(n_{j \downarrow}-\frac{1}{2}) .
\label{hamiltonian.hubbard}
\end{align}
is the most interesting model.
Here $c^{\dagger}_{j \sigma}$ and $c_{j \sigma}$  
are the fermion creation and annihilation operators 
and $n_{j \sigma}$ is the fermion number operator,
\begin{align} 
n_{j \sigma} = c^{\dagger}_{j \sigma} c_{j \sigma}.
\end{align} 
The parameter ${U}$ is the coupling constant 
describing the Coulomb interaction between electrons 
with opposite spins on the same site.
Lieb and Wu \cite{lieb} diagonalized the Hamiltonian 
(\ref{hamiltonian.hubbard}) by means of 
the coordinate Bethe ansatz method 
under the periodic boundary condition (PBC),
\begin{align}
 c^{\dagger}_{N+1, \sigma} = c^{\dagger}_{1 \sigma},
 \hspace{5mm}
 c_{N+1, \sigma} = c_{1 \sigma}, 
 \hspace{2cm}               \sigma = \uparrow \downarrow.
\label{pbcfermion}
\end{align}
The bulk properties of the 1D Hubbard model have been investigated 
by analyzing the associated Bethe ansatz equations. \cite{korepin}

The integrability of the 1D Hubbard model has been studied
by Shastry \cite{shastry1,shastry2,shastry3} 
and Wadati {\it et al.}. \cite{wadati1,wadati2,wadati3}
Shastry introduced the Jordan-Wigner transformation,\cite{jordan}
\begin{align}
   c_{j \uparrow} =& \exp \Bigl\{ 
         \mbox{i} \pi 
            \sum_{k=1}^{j-1} (n_{k \uparrow} - 1) \Bigr\}  
            \sigma_j^-,
\nonumber\\
   c_{j \downarrow} =& \exp \Bigl\{ 
         \mbox{i} \pi
            \sum_{k=1}^{j-1} (n_{k \downarrow} - 1) \Bigr\} 
                       \exp \Bigl\{ 
         \mbox{i} \pi
            \sum_{k=1}^{N} (n_{k \uparrow} - 1) \Bigr\}  
            \tau_j^- ,
\label{jwtransf}
\end{align}
to change the fermionic Hamiltonian (\ref{hamiltonian.hubbard})
into an equivalent coupled spin model,
\begin{align}
H = \sum_{j=1}^N (\sigma_{j+1}^+ \sigma_j^- + \sigma_j^+ 
                  \sigma_{j+1}^-)  
    + \sum_{j=1}^N (\tau_{j+1}^+ \tau_j^- + \tau_j^+ 
                    \tau_{j+1}^-)  
    + \dfrac{U}{4} \sum_{j=1}^N \sigma_j^z \tau_j^z, 
\label{coupledspin}
\end{align}
where ${\sigma}$  and ${\tau}$ are two species 
of the Pauli matrices commuting each other, and
\begin{align}
\sigma_j^{\pm} = \frac{1}{2} (\sigma_j^x \pm {\rm i} 
\sigma_j^y),
\hspace{5mm}
\tau_j^{\pm} = \frac{1}{2} (\tau_j^x \pm {\rm i} \tau_j^y).
\end{align}
Shastry constructed the $L$-operator and $R$-matrix 
which satisfy the Yang-Baxter relation,
\begin{align}
R_{12}(u_1,u_2) L_{a1}(u_1) L_{a2}(u_2)
 = L_{a2}(u_2) L_{a1}(u_1) R_{12}(u_1,u_2).
\label{spinyangr}
\end{align}
In ref. \cite{shastry3}, the decorated Yang-Baxter equation (DYBE)
for the free-fermion model was used to prove the Yang-Baxter relation
for the coupled spin model, (\ref{spinyangr}). It is a relation similar to, 
but not equivalent to the Yang-Baxter equation (YBE), 
Recently Shiroishi and Wadati \cite{shiroishi1,shiroishi2}
found that the YBE and the DYBE for the free-fermion model
can be generalized to a larger set of relations among 
the $L$-operators called tetrahedral Zamolodchikov algebra (TZA).
\cite{korepanov}
By use of the TZA, the Yang-Baxter equation for the $R$-matrix,  
\begin{align}
R_{12}(u_1,u_2) R_{13}(u_1,u_3) R_{23}(u_2,u_3)
= R_{23}(u_2,u_3) R_{13}(u_1,u_3) R_{12}(u_1,u_2)
\end{align}
can be proved. \cite{shiroishi1,shiroishi2,shiroishi3}
 
The coupled spin model (\ref{coupledspin}) 
is often referred to as the 1D Hubbard model,
since it is related to the fermionic Hamiltonian
(\ref{hamiltonian.hubbard}) through 
the Jordan-Wigner transformation (\ref{jwtransf}).
However, because of the non-locality of the Jordan-Wigner 
 transformation (\ref{jwtransf}), the PBC for the fermion model 
(\ref{pbcfermion}) does not correspond to the PBC for 
the coupled spin model.
Moreover the ${SO(4)}$ symmetry \cite{heilmann,yang1,yang2,pernici,affleck}, 
which is one of the most important properties 
of the Hamiltonian (\ref{hamiltonian.hubbard}),
is not clear in the coupled spin model
(\ref{coupledspin}). 

To discuss the fermionic Hamiltonian directly,
we can employ the fermionic formulation of 
the Yang-Baxter relation 
(graded Yang-Baxter relation). \cite{wadati2} 
In this formulation, the transfer matrix 
that respects the PBC (\ref{pbcfermion}) can be  
constructed by taking the supertrace of the monodromy matrix. 
Recently it has been shown that the transfer matrix 
naturally enjoys the ${SO(4)}$ symmetry.
\cite{murakami,shiroishi4} 
Based on this fermionic formulation, 
the algebraic Bethe ansatz method has been developed 
by Ramos and Martins. \cite{ramos,martins}  

The aim of this paper is to present yet another approach 
to discuss the integrability of the 1D Hubbard model 
using the fermionic ${R}$-operator. 
The notion of the fermionic ${R}$-operator was introduced 
in relation to the ${XXZ}$ fermion model.\cite{destri,umeno} 
The fermionic ${R}$-operator consists of the fermion operators 
and is a solution of the YBE.  

In this paper, we first consider the ${XYZ}$ fermion model. 
The fermionic ${R}$-operator for the ${XYZ}$ fermion model, 
which satisfies the YBE, is explicitly obtained. The YBE ensures 
the commutativity of the transfer operators. The logarithmic 
derivative of the transfer operator yields the Hamiltonian of 
the ${XYZ}$ fermion model under the PBC. 

Next we consider the DYBE for the fermionic ${R}$-operator 
and find that the fermionic ${R}$-operator related to 
the ${XY}$ fermion  model fulfills the DYBE. 
Combining the YBE and the DYBE of the fermionic ${R}$-operator 
for the ${XX}$ fermion models,
we construct a new fermionic ${R}$-operator and an ${L}$-operator 
for the 1D Hubbard model. The transfer operator constructed 
from the ${L}$-operator generates the Hamiltonian and 
other conserved operators under the PBC. 

We also prove the YBE 
for the new integrable ${R}$-operator 
by considering the fermionic variant of 
the tetrahedral Zamolodchikov algebra (TZA).  
Then we consider the generalized transfer operator which is 
constructed from the fermionic ${R}$-operator.  
A new fermionic Hamiltonian is obtained from the transfer 
operator.

It is possible to discuss the ${SO(4)}$ symmetry of 
the 1D Hubbard model using the fermionic ${R}$-operator. 
The (anti) commutation relations between the fermionic ${R}$-operator 
and the generators of the ${SO(4)}$ algebra are obtained. 
Those relations are extended to the  monodromy operator and 
the transfer operator. We can show the ${SO(4)}$ invariance 
of the local conserved operators, 
though the transfer operator itself is not ${SO(4)}$ invariant. 

The outline of this paper is as follows. 
In \S 2, we study the Yang-Baxter equation (YBE) 
for the ${XYZ}$ fermion model. 
Fermionic ${R}$-operator related to the ${XYZ}$ fermion model is 
explicitly shown by use of Baxter's parameterization. 
In \S 3, the decorated Yang-Baxter equation (DYBE) 
for the fermionic ${R}$-operator is introduced. 
It is necessary to impose the free-fermion condition 
on the Boltzmann weights in order that the DYBE is fulfilled. 
Some connections to the ${XY}$ fermion model 
with an external field are also discussed. 
In \S 4, we discuss the integrability of the 1D Hubbard model. 
The fermionic ${R}$-operator for the 1D Hubbard model 
is constructed from  the ${R}$-operators
related to the ${XX}$ fermion models with spin up and down.
The YBE and the DYBE are utilized to show the Yang-Baxter relation,
where a new ${L}$-operator is also introduced.
 In \S 5, we prove the YBE
for the fermionic ${R}$-operator of the Hubbard model. 
We use the fermionic variant of the TZA. The discussions proceed
just in parallel with the case of the coupled spin model 
(\ref{coupledspin}). We also use a generalized transfer 
operator which consists of the fermionic ${R}$-operators.
In \S 6, we give a new approach 
to the ${SO(4)}$ symmetry of 
the 1D Hubbard model. The fermionic ${R}$-operator has a natural
${SO(4)}$ symmetry. We show that all the local fermionic 
conserved operators derived from the transfer operator 
are ${SO(4)}$ invariant.
The last section is devoted to the concluding remarks.

%%%%%%%%%%%%%%%%%%%%%%%%%%%%%%%%%%%%%%%%%%%%%%%%%%%%%%%%%%%%%%%%%%%%%%

\section{Fermionic ${R}$-Operator for the ${XYZ}$ Fermion Model}
\setcounter{equation}{0}
\renewcommand{\theequation}{2.\arabic{equation}}
It is well understood that the Yang-Baxter equation (YBE)
\begin{align}
\mathcal{R}_{12}(u-v) \mathcal{R}_{13}(u) \mathcal{R}_{23}(v)
= \mathcal{R}_{23}(v) \mathcal{R}_{13}(u) \mathcal{R}_{12}(u-v),
\label{str}
\end{align}
plays the most important role in the quantum integrable models.
\cite{baxter,kulish,sklyanin,korepin2,gomez}

We look for a solution of the YBE (\ref{str}) in the form 
\begin{align}
       \mathcal{R}_{jk}(u) =& a(u) \{- n_j n_k + (1-n_j)(1-n_k)\}
           + b(u) \{n_j (1-n_k) + (1-n_j) n_k\}
\nonumber\\    
           +  & c(u) (c_j^{\dagger} c_k + c_k^{\dagger} c_j)
           - d(u) (c_j^{\dagger} c_k^{\dagger} + c_j c_k),
\label{R}
\end{align}
where ${c_{j}^{\dagger}}$ and ${c_{j}}$ are the fermion creation 
annihilation operators satisfying the canonical anti-commutation 
relations
\begin{align}
\big\{ c_{j}, c_{k} \big\} = 
\big\{ c_{j}^{\dagger}, c_{k}^{\dagger} \big\} = 0,\hspace{1cm}
\big\{ c_{j}^{\dagger}, c_{k} \big\} = \delta_{jk},
\label{anticommu}
\end{align}
and ${n_{j} = c_{j}^{\dagger} c_{j}}$. 

Substituting (\ref{R}) into (\ref{str}) and using (\ref{anticommu}), 
we get the following relations among the functions ${a(u)}$, 
${b(u)}$, ${c(u)}$ and
${d(u)}$,
%The relation of $a(u)$, $b(u)$, $c(u)$ and $d(u)$ 
\begin{align}
a(u-v)c(u)a(v) + d(u-v)a(u)d(v) &= b(u-v)c(u)b(v) + c(u-v)a(u)c(v),
\nonumber\\
a(u-v)b(u)c(v) + d(u-v)d(u)b(v) &= b(u-v)a(u)c(v) + c(u-v)c(u)b(v),
\nonumber\\
c(u-v)b(u)a(v) + b(u-v)d(u)d(v) &= c(u-v)a(u)b(v) + b(u-v)c(u)c(v),
\nonumber\\
a(u-v)d(u)b(v) + d(u-v)b(u)c(v) &= b(u-v)d(u)a(v) + c(u-v)b(u)d(v),
\nonumber\\
a(u-v)a(u)d(v) + d(u-v)c(u)a(v) &= b(u-v)b(u)d(v) + c(u-v)d(u)a(v),
\nonumber\\
d(u-v)a(u)a(v) + a(u-v)c(u)d(v) &= d(u-v)b(u)b(v) + a(u-v)d(u)c(v).
\label{baxterr}
\end{align}
These are nothing but the relations among the Boltzmann weights of the 8-vertex model (Baxter model). \cite{baxter} Then, using Baxter's parameterization, 
we can solve (\ref{baxterr}) as follows,
\begin{align}
a(u) &= \sn (u+2\eta),       
\hspace{5mm}
b(u) = \sn (u),
\hspace{5mm}
c(u) = \sn (2\eta),
\nonumber\\
d(u) &= k a(u) b(u) c(u) = k \sn (2\eta) \sn (u) \sn (u+2\eta),
\label{baxterpara}
\end{align}
where ${\sn u= \sn (u,k)}$ is Jacobi's elliptic function with the modulus ${k}$ and ${\eta}$ is the anisotropy parameter.

Thus we have found a new type of solution for the YBE (\ref{str}). 
We call (\ref{R}) the fermionic ${R}$-operator (related 
to the ${XYZ}$ fermion model). We remark that Destri and Segalini
\cite{destri} have discussed the special case (${k=0}$) of (\ref{R}).   
The fermionic $R$-operator (\ref{R}) can also be obtained from the 
${R}$-matrix of the ${XYZ}$ spin model (Appendix A).

It is remarkable that $\mathcal{R}_{jk}(u)$ has a property,
\begin{align}
\mathcal{R}_{jk}(0) = \sn 2 \eta \mathcal{P}_{jk},
\label{regularity}
\end{align}
where ${\mathcal{P}_{jk}}$ is the permutation operator,
\begin{align}
&\mathcal{P}_{jk} = 1 - (c_{j}^{\dagger} - c_{k}^{\dagger})
(c_{j} - c_{k}) = 1 - n_{j} - n_{k} + c_{j}^{\dagger} c_{k}
+ c_{k}^{\dagger} c_{j}, \nonumber \\
&\mathcal{P}_{jk} c_j = c_k \mathcal{P}_{jk}, \ \ \ \ 
\mathcal{P}_{jk} c_j^{\dagger} = c_k^{\dagger} \mathcal{P}_{jk}.
\label{fermionicpermutation}
\end{align}
This property is useful when we derive a related Hamiltonian
from the transfer operator.

Unlike the ${R}$-matrix of the usual ${XYZ}$ spin model,
the fermionic ${R}$-operator is not symmetric, ${\mathcal{R}_{12}(u) 
\ne \mathcal{R}_{21}(u)}$. Instead, the following relation holds
\begin{align}
&\mathcal{R}_{12}(u;k) = \mathcal{R}_{21}(u;-k),
\label{modulusinverse}
\end{align}
where we write the modulus ${k}$ dependence explicitly.
The fermionic ${R}$-operator (\ref{R}) also has the unitarity 
relation in the form
\begin{align}
\mathcal{R}_{12}(u) \mathcal{R}_{21}(-u) = 
\left( - \sn^{2} u + \sn^{2} 2 \eta \right) {\bf 1},
\end{align}
as well as the property
\begin{align}
\left[ \mathcal{R}_{jk}(u), (1 - 2 n_{j})(1 - 2 n_{k}) \right] = 0.
\label{generalproperty}
\end{align}

Now we introduce the monodromy operator 
\begin{align}
\mathcal{T}_{a}(u) = \mathcal{R}_{aN}(u) \ldots \mathcal{R}_{a1}(u),
\label{monodromy}
\end{align}
where the suffix {\it a} means the auxiliary space,
The transfer operator is given by the ``supertrace" of the monodromy
operator 
\begin{align}
\tau(u) &= \fstr_{a} \mathcal{T}_{a}(u).
\label{transfer}
\end{align}
Here we define the supertrace  ${\fstr}$ as
\begin{align} 
\fstr_{a} X = {}_{a} \langle 0 | X | 0 \rangle_{a}
         - {}_{a} \langle 1 | X | 1 \rangle_{a},
\label{fstr}
\end{align}
with the auxiliary fermion Fock space being
\begin{align}
c_{a} | 0 \rangle_{a} = 0, 
\hspace{5mm} 
| 1 \rangle_{a} = c_{a}^{\dagger} | 0 \rangle_{a}, 
\hspace{5mm} {}_a \langle 0 | = ( | 0 \rangle_{a})^{\dagger}, 
\hspace{5mm} {}_{a} \langle 1 | = {}_{a} \langle 0 | c_{a},
\end{align}
\begin{align}
{}_a \langle  0 | 0 \rangle_a
= {}_a \langle  1 | 1 \rangle_a=1.
\end{align}
When we write the monodromy operator ${\mathcal{T}_{a}(u)}$ as
\begin{align}
\mathcal{T}_{a}(u) = A(u) (1 - n_{a}) + B(u) c_{a} + C(u) c_{a}^{\dagger}
+ D(u) n_{a},
\label{ABCD}
\end{align}
 the transfer operator (\ref{transfer}) is given by
\begin{align}
\tau(u) = A(u) - D(u).
\label{AD}
\end{align}

The YBE for the fermionic ${R}$-operator leads to the global 
Yang-Baxter relation for the monodromy operator
(\ref{monodromy}),
\begin{align}
             \mathcal{R}_{12}(u-v)
     \mathcal{T}_1(u) \mathcal{T}_2(v)
           &=      \mathcal{T}_2(v) \mathcal{T}_1(u)
     \mathcal{R}_{12}(u-v).
\label{gyangrxyz}
\end{align}
Rewriting (\ref{gyangrxyz}) as
\begin{align}
             \mathcal{R}_{12}(u-v)
     \mathcal{T}_1(u) \mathcal{T}_2(v) \mathcal{R}_{12}^{-1} (u-v)
           &=      \mathcal{T}_2(v) \mathcal{T}_1(u),
\end{align}
and taking the supertrace (\ref{fstr}) of both sides,  
we can show the commutativity
of the transfer operators
\begin{align}
\left[ \tau(u), \tau(v) \right] = 0,
\label{trans}
\end{align}
which ensures the existence of enough number of the fermionic 
conserved operators. 
In this way we can establish the integrability of the 
fermion model by means of the fermionic ${R}$-operator. 

Using (\ref{regularity}) in (\ref{transfer}) with (\ref{monodromy}), we have
\begin{align}
\tau(0) &= \left( \sn 2 \eta \right)^{N} \ \fstr_{a} (\mathcal{P}_{aN} \ldots
                      \mathcal{P}_{a2} \mathcal{P}_{a1}) \nonumber \\
        &= \left( \sn 2 \eta \right)^{N} \ \mathcal{P}_{12} \mathcal{P}_{23} 
                          \ldots \mathcal{P}_{N-1,N}, 
\end{align}
where we have used some fundamental properties of the permutation operators  
\begin{align}
&\mathcal{P}_{aj} \mathcal{P}_{ak} = \mathcal{P}_{jk} \mathcal{P}_{aj},
\hspace{1cm} \mathcal{P}_{jk} = \mathcal{P}_{kj}, \nonumber \\
&\fstr_{a} \mathcal{P}_{aj} = 1.
\end{align}
Thus we find that ${\tau(0)}$ is proportional to the (left) shift 
operator ${\hat{U}}$ for fermions,
\begin{align}
\hat{U} = \mathcal{P}_{12} \mathcal{P}_{23} 
                          \ldots \mathcal{P}_{N-1,N} .
\end{align}
It is easy to confirm that the shift operator ${\hat{U}}$ 
acts on the fermion operators as 
\begin{align}
&\hat{U} c_{j} = c_{j+1} \hat{U}, \hspace{3cm} j=1, \ldots, N-1,
\nonumber \\
&\hat{U} c_{N} = c_{1} \hat{U}.
\end{align}

In a similar way as the spin models, the Hamiltonian of the ${XYZ}$ 
fermion model is derived from the logarithmic derivative of 
the transfer operator
\begin{align}
\mathcal{H}^{\rm XYZ} &=  \tau(0)^{-1} \frac{\rm d}{{\rm d} u} 
\tau(u) \Big|_{u=0}  \nonumber \\
&= \sum_{j=1}^N  \mathcal{H}_{j,j+1}^{\rm XYZ},
\label{xyz}
\end{align}
where the two-point Hamiltonian density is given by
\begin{align}
\mathcal{H}_{j,j+1}^{\rm XYZ} 
     =& \frac{1}{\sn 2 \eta} \Big\{ a'(0) \left( n_{j} n_{j+1} 
                                 + (1-n_{j})(1-n_{j+1}) \right)
\nonumber\\   
     & \hspace{5mm} + b'(0) ( c_{j}^{\dagger} c_{j+1} + 
                              c_{j+1}^{\dagger} c_{j} )
     + d'(0) ( c_{j}^{\dagger} c_{j+1}^{\dagger} 
               - c_{j} c_{j+1}) \Big\}.   
\label{XYZhamiltonian}
\end{align}
Here and hereafter, ${'}$ means the derivative with respect to the spectral
parameter ${u}$, 
\begin{align}
a'(0) = \cn 2 \eta \ \dn 2 \eta, 
\hspace{5mm} b'(0) = 1, \hspace{5mm}
d'(0) = k \ \sn^{2} 2 \eta.
\end{align}
In (\ref{xyz}), we have used a fact that $\tau(0)^{-1}$
is proportional to the shift operator.
In terms of an equivalent ${R}$-operator 
${\check{\mathcal{R}}_{jk}(u) \equiv \mathcal{P}_{jk} %
\mathcal{R}_{jk}(u)}$ (c.f. Appendix A), 
the two-point Hamiltonian density is expressed as
\begin{align}
\mathcal{H}_{jk}^{\rm XYZ} = \dfrac{\rm d}{{\rm d} u} 
                             \check{R}_{jk}(u) \Big|_{u=0}.
\end{align}

One may wonder if  our discussion using the fermionic ${R}$-operator 
goes fully in parallel with the standard treatment of the integrable 
spin models and no new aspect appears. In fact we can proceed the discussion almost in the same way as the spin models. However we like to point out an important difference between 
the spin model and the fermion model here. 
While the Hamiltonian (\ref{XYZhamiltonian}) is hermitian,
it is not invariant under the space inversion
\begin{align}
j \longrightarrow N-j+1.
\label{inversion}
\end{align}
It is easy to see that the modulus ${k}$ 
in the Hamiltonian (\ref{XYZhamiltonian}) 
changes its sign under the transformation (\ref{inversion}).
This point makes a difference between the ${XYZ}$ spin model 
and the ${XYZ}$ fermion model. Actually it is a consequence of 
the property (\ref{modulusinverse}).

In the case of $k=0$, the fermionic ${R}$-operator (\ref{R}) 
reduces to the one with 
\begin{align}
a(u) &= \sin (u+2\eta)       
\hspace{5mm}
b(u) = \sin u
\hspace{5mm}
c(u) = \sin 2\eta
\hspace{5mm}
d(u) = 0,
\end{align}
which we have used in ref. \cite{umeno} to discuss
 the integrability of the ${XXZ}$ fermion model
\begin{align}
\mathcal{H}^{\rm XXZ} 
     =& \frac{1}{\sin 2 \eta} \sum_{j=1}^{N} 
 \Big\{  c_{j}^{\dagger} c_{j+1} + c_{j+1}^{\dagger} c_{j} 
        + \cos 2 \eta \left( n_{j} n_{j+1} + (1-n_{j})(1-n_{j+1}) 
\right) \Big\}.
\label{XXZHamiltonian} 
\end{align}
In this case, the fermionic ${R}$-operator becomes symmetric 
and the Hamiltonian
 (\ref{XXZHamiltonian}) is invariant under 
the space inversion (\ref{inversion}).

We note that the periodic boundary condition (PBC) 
for the fermion operators is satisfied in (\ref{XYZhamiltonian}),
\begin{align}
c_{N+1} = c_{1}, \hspace{1cm} c_{N+1}^{\dagger} = c_{1}^{\dagger}.
\end{align}
It is sometimes necessary to consider the anti-periodic 
boundary condition (APBC) for the fermion operators
\begin{align}
c_{N+1} = - c_{1}, \hspace{1cm} c_{N+1}^{\dagger} = - c_{1}^{\dagger}.
\label{apbc}
\end{align}
To explain this, we consider the following twisted monodromy operator 
\begin{align}
\mathcal{\tilde{T}}_{a}(u) &= (1 - 2 n_{a}) \mathcal{T}_{a}(u) \nonumber \\
                           &= (1 - 2 n_{a}) \mathcal{R}_{aN}(u) 
\ldots \mathcal{R}_{a1}(u).
\label{twistedmonodromy}
\end{align}
Then, thanks to the property (\ref{generalproperty}), 
the global Yang-Baxter relation 
for the twisted monodromy operator
 holds,
\begin{align}
             \mathcal{R}_{12}(u-v)
     \mathcal{\tilde{T}}_1(u) \mathcal{\tilde{T}}_2(v) 
           &= \mathcal{\tilde{T}}_2(v) \mathcal{\tilde{T}}_1(u) 
\mathcal{R}_{12}(u-v),
\end{align}
which leads to the commutativity of 
the twisted transfer operator ${\tilde{\tau}(u) = \fstr_{a} 
\mathcal{\tilde{T}}_a(u)}$,
\begin{align}
\left[ \tilde{\tau}(u), \tilde{\tau}(v) \right] = 0.
\end{align}
We can show that the twisted transfer operator generates 
the Hamiltonian (\ref{XYZhamiltonian}) under the APBC 
(\ref{apbc}). In fact the logarithmic derivative of  
${\tilde{\tau}(u)}$ yields a Hamiltonian
\begin{align}
\mathcal{\tilde{H}}^{\rm XYZ} 
= & \tilde{\tau}(0)^{-1} \frac{\rm d}{{\rm d} u} 
\tilde{\tau}(u) \Big|_{u=0}  \nonumber \\
= & \sum_{j=1}^{N-1}  \mathcal{H}_{j,j+1}^{\rm XYZ} 
    + \frac{1}{\sn 2 \eta} \Big\{ a'(0) \left( n_{N} n_{1} 
                                 + (1-n_{N})(1-n_{1}) \right)
\nonumber\\   
     & \hspace{3.5cm} - b'(0) ( c_{N}^{\dagger} c_{1} + 
                              c_{1}^{\dagger} c_{N} )
     - d'(0) ( c_{N}^{\dagger} c_{1}^{\dagger} 
               - c_{N} c_{1}) \Big\}.  
\end{align}
Here we have used a relation
\begin{align}
(1 - 2 n_{N}) \fstr_{a} \left\{ (1 - 2 n_{a}) \mathcal{P}_{aN} 
\mathcal{H}_{a1}^{XYZ} \right\} 
= (1 - 2 n_{N}) \mathcal{H}_{N1}^{XYZ} (1 - 2 n_{N}),
\end{align}
as well as the identities
\begin{align}
( 1 - 2 n_{N}) c_{N} (1 - 2 n_{N}) &= - c_{N}, \hspace{5mm}
( 1 - 2 n_{N}) c_{N}^{\dagger} (1 - 2 n_{N}) = - c_{N}^{\dagger}.
\end{align} 
Note that if we express ${\mathcal{T}_a(u)}$ as (\ref{ABCD}), 
the twisted transfer operator takes the following form, 
\cite{destri}
\begin{align}
\tilde{\tau}(u) = A(u) + D(u).
\end{align}

%%%%%%%%%%%%%%%%%%%%%%%%%%%%%%%%%%%%%%%%%%%%%%%%%%%%%%

\section{Decorated Yang-Baxter Equation}
\setcounter{equation}{0}
\renewcommand{\theequation}{3.\arabic{equation}}
%In ref. \cite{shastry3}, Shastry introduced 
The decorated Yang-Baxter equation (DYBE) is an algebraic relation 
similar to the YBE (\ref{str}), but not equivalent to the YBE. 
Shastry \cite{shastry3} showed that the free-fermion condition on 
the Boltzmann weights is necessary 
so that the solution of the YBE also satisfies 
the DYBE. \cite{shastry3,shiroishi2} 
The DYBE is considered to be a hidden algebraic structure of 
the free-fermion model (equivalently, ${XY}$ model).

We discuss the DYBE in the context of the fermionic $R$-operator. 
The DYBE for the fermionic ${R}$-operator is defined by
\begin{align}
\mathcal{R}_{12}(u+v) (2n_1-1) \mathcal{R}_{13}(u) \mathcal{R}_{23}(v)
= \mathcal{R}_{23}(v) \mathcal{R}_{13}(u) (2n_1-1) \mathcal{R}_{12}(u+v).
\label{dstr}
\end{align}
The DYBE (\ref{dstr}) looks alike the YBE (\ref{str}), but the 
operator ${(2n_{1}-1)}$ inserted in both sides 
can not be absorbed by the redefinition of ${R_{jk}(u)}$.
It is possible to simplify the DYBE by use of the YBE as follows.

First, 
by multiplying ${(2n_1-1)(2n_2-1)}$ to the 
YBE (\ref{str}) 
from the left and $(2n_3-1)$ from the right,
we get a relation 
\begin{align}
&\mathcal{R}_{12}(u-v) (2n_1-1) \mathcal{R}_{13}(u) 
(2n_2-1) \mathcal{R}_{23}(v) (2n_3-1) \nonumber \\
&= (2n_2-1) \mathcal{R}_{23}(v) (2n_3-1) 
\mathcal{R}_{13}(u) (2n_1-1) \mathcal{R}_{12}(u-v),
\label{modSTR}
\end{align}
where we have used a general property (\ref{generalproperty})
and an equivalent relation,
\begin{align}
(2n_j-1) \mathcal{R}_{jk}(u) (2n_k-1) = 
(2n_k-1) \mathcal{R}_{jk}(u) (2n_j-1).
\end{align}
Then comparing eq.\ (\ref{modSTR}) with the definition of 
the DYBE (\ref{dstr}), 
we find a condition
\begin{align}
(2n_j-1) &\mathcal{R}_{jk}(u) (2n_k-1)
    = \mathcal{R}_{jk}(-u).
\label{condition}
\end{align}
Hence we see that the DYBE is equivalent to the condition (\ref{condition}) 
when we have the YBE.
In terms of the Boltzmann weights,
the condition (\ref{condition}) is rephrased as 
\begin{align}
a(u) = a(-u), \hspace{5mm} -b(u) = b(-u), \hspace{5mm} 
c(u) = c(-u), \hspace{5mm} -d(u) = d(-u).
\label{condition2}
\end{align}
Using the explicit parameterization (\ref{baxterpara}), 
we find that the condition (\ref{condition2}) corresponds to 
\begin{align}
2 \eta = K,\ \  3K,
\end{align}
where ${K}$ is the complete elliptic 
integral of the first kind. 
In this case, the free-fermion condition 
\begin{align}
a^2(u) + b^2(u) = c^2(u) + d^2(u),
\label{FFcond}
\end{align}
is fulfilled. Thus as in the case of the spin models, 
we have shown that the free-fermion condition is required
for the validity of the DYBE (\ref{dstr}). 

Hereafter we restrict our consideration to the case 
${2 \eta = K}$. 
Then, from (\ref{baxterpara}), the Boltzmann weights 
${a(u),b(u),c(u)}$ 
and ${d(u)}$ are explicitly given as
\begin{align}
&a(u) = \cd u, \hspace{5mm} b(u)= \sn u, 
\hspace{5mm} c(u)=1, 
\nonumber \\
&d(u) = k \cd u \ \sn u, 
\label{XYBoltzmann}
\end{align}
where ${\cd u = \cn u / \dn u}$.
The corresponding Hamiltonian is the ${XY}$ fermion model
\begin{align}
\mathcal{H}^{\rm XY} 
= \sum_{j=1}^{N} \Big\{ 
c_{j}^{\dagger} c_{j+1} + c_{j+1}^{\dagger} c_{j}
+ k \left( c_{j}^{\dagger} c_{j+1}^{\dagger} -
 c_{j} c_{j+1} \right) \Big\}.   
\label{XYhamiltonian}
\end{align}
In the special case $k=0$, this Hamiltonian reduces 
to the $XX$ fermion model
\begin{align}
\mathcal{H}^{\rm XX} 
     = \sum_{j=1}^{N} ( c_{j}^{\dagger} c_{j+1} + 
c_{j+1}^{\dagger} c_{j}).   
\label{XXhamiltonian}
\end{align}

As an interesting application of the YBE (\ref{str}) 
and the DYBE (\ref{dstr}), 
we shall prove the integrability of the ${XY}$ fermion model 
in an external field,
\begin{align}
\mathcal{H}^{\rm XY \mu} = \sum_{j=1}^{N} 
  \Big\{ c_{j}^{\dagger} c_{j+1} + c_{j+1}^{\dagger} c_{j}
  + k \left( c_{j}^{\dagger} c_{j+1}^{\dagger} - c_{j} c_{j+1} \right) 
+ \mu \left( n_{j} - \frac{1}{2} \right) \Big\}.
\label{XYfield}
\end{align}
The external field ${\mu}$ can be regarded as the chemical 
potential for the ${XY}$ fermion model. 
Since the term 
${\sum_{j=1}^{N} (n_{j} - 1/2)}$ does not commute with 
the Hamiltonian (\ref{XYhamiltonian}) unless ${k=0}$, 
the addition of the extra term is not trivial at all. 
It is well known that the model (\ref{XYfield}) can be 
diagonalized 
by means of the Fourier transformation through the Bogoliubov
transformation.
 \cite{lieb2} The case ${k=\pm 1}$ 
corresponds to the transverse Ising model. \cite{pfeuty}  

The following argument is in common with the one 
explored for the spin models. \cite{shastry1,shiroishi2} 
From now on, let  ${\mathcal{R}_{jk}(u)}$ denote 
the fermionic ${R}$-operator 
for the ${XY}$ fermion model. 
Taking a linear combination of the YBE (\ref{str}) and 
the DYBE (\ref{dstr}), we have
\begin{align}
& \Big\{ \alpha \mathcal{R}_{12}(u-v) 
  + \beta \mathcal{R}_{12}(u+v)(2n_1-1) \Big\}
     \mathcal{R}_{13}(u) \mathcal{R}_{23}(v)
\nonumber\\
&= \mathcal{R}_{23}(v) \mathcal{R}_{13}(u) 
     \Big\{ \alpha \mathcal{R}_{12}(u-v) 
           + \beta (2n_1-1) \mathcal{R}_{12}(u+v) \Big\},
\label{com}
\end{align}
where ${\alpha}$ and ${\beta}$ are not operators but constants.
 
Now we look for a solution of the  Yang-Baxter relation 
\begin{align}
         \mathcal{R}^{\mu}_{12}(u,v) 
         \mathcal{L}_{13} (u)
         \mathcal{L}_{23} (v) 
    &=    \mathcal{L}_{23} (v)
         \mathcal{L}_{13} (u)
         \mathcal{R}^{\mu}_{12} (u,v),
\label{yang}
\end{align}
assuming the ${R}$-operator ${\mathcal{R}^{\mu}_{12}(u,v)}$
and the ${L}$-operator ${\mathcal{L}_{jk} (u)}$ in the form, 
\begin{align}
&\mathcal{R}^{\mu}_{12}(u,v)  = \alpha \mathcal{R}_{12}(u-v) 
  + \beta \mathcal{R}_{12}(u+v)(2n_1-1), \label{Rxy} \\
&\mathcal{L}_{jk}(u) = \mathcal{R}_{jk}(u) 
\exp \left\{ h(u) ( 2n_{j}-1) 
\right\}.
\label{Lxy}
\end{align}
Here ${\alpha,\beta}$ and ${h(u)}$ are functions 
of the spectral parameters to be specified later. 

Comparing eq.\ (\ref{yang}) with eq.\ (\ref{com}),
we get a relation
\begin{align}
&\mathcal{I}_1(u) \mathcal{I}_2(v)
 \mathcal{R}^{\mu}_{12}(u,v) \mathcal{I}_1(u)^{-1} 
  \mathcal{I}_2(v)^{-1}
 = \alpha \mathcal{R}_{12}(u-v) + 
   \beta (2n_1-1)\mathcal{R}_{12}(u+v),
\label{condxy}
\end{align}
where
\begin{align}
\mathcal{I}_j(u) = \exp\{h(u) (2n_j-1)\}.
\end{align}

For $k \neq 0$ (that is, the genuine $XY$ fermion model), the 
relation (\ref{condxy}) gives two conditions,
\begin{align}
  \dfrac{\beta}{\alpha} = \tanh (h(u)-h(v))
\label{ratio1}
\\
 \dfrac{\beta d(u+v)}{\alpha d(u-v)} = \tanh (h(v)+h(u)).
\label{ratio2}
\end{align}
Equations (\ref{ratio1}) and (\ref{ratio2}) give the ratio of 
$\alpha$ to $\beta$
and constraints on $h(u)$ and ${h(v)}$. Substituting 
(\ref{XYBoltzmann}) into (\ref{ratio2}), 
we find that the constraints are given by
\begin{align}
\frac{\sinh 2 h(u)}{ \bsc 2 u} = 
\frac{\sinh 2 h(v)}{ \bsc 2 v}
= {\rm constant},
\label{sc2}
\end{align}
where ${\bsc u = \sn u / \cn u}$.  
In conclusion, we have proved that the fermionic ${R}$-operator
\begin{align}
\mathcal{R}_{12}^{\mu}(u,v) = \mathcal{R}_{12}(u-v) + 
\tanh (h(u) - h(v)) \mathcal{R}_{12}(u+v) (2n_1-1),
\end{align}
satisfies the Yang-Baxter relation (\ref{yang}) with 
the ${L}$-operators 
(\ref{Lxy}) under the constraints (\ref{sc2}) . 

From eq.\ (\ref{yang}) we get the global Yang-Baxter relation for 
the monodromy operator 
\begin{align}
\mathcal{T}_{a}(u) = 
\mathcal{L}_{aN}(u) \ldots \mathcal{L}_{a1}(u),
\end{align}
as
\begin{align}
             \mathcal{R}^{\mu}_{12}(u,v)       
     \mathcal{T}_1(u) \mathcal{T}_2(v)
           =     \mathcal{T}_2(v) \mathcal{T}_1(u)
     \mathcal{R}^{\mu}_{12}(u,v),
\end{align}
which leads to a one-parameter family of the commuting 
transfer operators,
\begin{align}
&  [ \tau(u) , \tau(v)] = 0, \nonumber \\
&\tau(u) = \fstr_a \mathcal{T}_a(u).
\end{align}

If we identify the constant in eq.\ (\ref{sc2}) with ${\mu/2}$, 
we recover the Hamiltonian (\ref{XYfield}) by
the logarithmic derivative of the transfer operator
\begin{align}
\mathcal{H} =& \tau(0)^{-1} \frac{{\rm d}}{{\rm d} u} \tau(u) 
\Big|_{u=0} 
\nonumber\\
            =&  \mathcal{H}^{\rm XY}  + 
\mu \sum_{j=1}^{N} \left( n_j-\frac{1}{2} 
\right).
\end{align}
Here we have used the relations ${h(0) = 0, \ \ h'(0) = \mu/2}$.

%%%%%%%%%%%%%%%%%%%%%%%%%%%%%%%%%%%%%%%%%%%%%%%%%%%%%%%%%%%%

\section{Fermionic $R$-operator for the 1D Hubbard model}
\setcounter{equation}{0}
\renewcommand{\theequation}{4.\arabic{equation}}
The Hamiltonian of the 1D Hubbard model 
(\ref{hamiltonian.hubbard}) consists of two $XX$ fermion models 
(\ref{XXhamiltonian}) for up and down spins with
an interaction term between them. 
The fermionic $R$-operator without an interaction term is 
given by
\begin{align}
\bar{\mathcal{R}}_{jk}(u) 
= \mathcal{R}^{(\uparrow)}_{jk}(u) 
\mathcal{R}^{(\downarrow)}_{jk}(u),
\end{align}
where ${\mathcal{R}_{jk}^{(\sigma)} \ (\sigma = \uparrow \downarrow)}$ denote the fermionic ${R}$-operators for the ${XX}$ fermion model, 
i.e.,
\begin{align}
a(u) = \cos u, \hspace{5mm} 
b(u) = \sin u, \hspace{5mm} 
c(u)=1, \hspace{5mm}
d(u) =0.
\end{align}
Since both ${\mathcal{R}^{(\uparrow)}_{jk}(u)}$ and 
${\mathcal{R}^{(\downarrow)}_{jk}(u)}$
satisfy the YBE and the DYBE, 
the product ${\bar{\mathcal{R}}_{jk}(u)}$ also satisfies the YBE 
\begin{align}
\bar{\mathcal{R}}_{12}(u-v) &\bar{\mathcal{R}}_{13}(u)
 \bar{\mathcal{R}}_{23}(v)
= \bar{\mathcal{R}}_{23}(v)
 \bar{\mathcal{R}}_{13}(u) \bar{\mathcal{R}}_{12}(u-v)
\label{hstr}
\end{align}
and the DYBE
\begin{align}
\bar{\mathcal{R}}_{12}(u+v) 
&(2n_{1 \uparrow}-1) (2n_{1 \downarrow}-1) 
\bar{\mathcal{R}}_{13}(u) \bar{\mathcal{R}}_{23}(v)
\nonumber\\
&= \bar{\mathcal{R}}_{23}(v) \bar{\mathcal{R}}_{13}(u) 
(2n_{1 \uparrow}-1) (2n_{1 \downarrow}-1) 
\bar{\mathcal{R}}_{12}(u+v).
\label{hdstr}
\end{align}
A linear combination of (\ref{hstr}) and (\ref{hdstr}) yields
\begin{align}
&\Bigl\{ \alpha \bar{\mathcal{R}}_{12}(u-v) 
     + \beta \bar{\mathcal{R}}_{12}(u+v)
        (2n_{1 \uparrow}-1)(2n_{1 \downarrow}-1)
            \Bigr\}
     \bar{\mathcal{R}}_{13}(u) \bar{\mathcal{R}}_{23}(v)
\nonumber\\
&= \bar{\mathcal{R}}_{23}(v) \bar{\mathcal{R}}_{13}(u) 
     \Bigl\{ \alpha \bar{\mathcal{R}}_{12}(u-v) 
         + \beta (2n_{1 \uparrow}-1)(2n_{1 \downarrow}-1) 
             \bar{\mathcal{R}}_{12}(u+v)
                \Bigr\}.
\label{hcom}
\end{align}
For a moment, $\alpha$ and $\beta$ are arbitrary.
As in the case of the ${XY}$ fermion model in an external field, 
we look for a solution of the Yang-Baxter relation,
\begin{align}
         \mathcal{R}^{\rm h}_{12}(u,v) 
            \mathcal{L}_{13}(u)
            \mathcal{L}_{23}(v)
&=       \mathcal{L}_{23}(v)
            \mathcal{L}_{13}(u)
         \mathcal{R}^{\rm h}_{12}(u,v) 
\label{YBRhubbard}
\end{align}
in the form
\begin{align}
\mathcal{R}^{\rm h}_{12}(u,v)  &=
\alpha \bar{\mathcal{R}}_{12}(u-v) 
+ \beta \bar{\mathcal{R}}_{12}(u+v)
(2n_{1 \uparrow}-1)(2n_{1 \downarrow}-1), \label{hubbardR} \\
   \mathcal{L}_{jk}(u)
       =& \bar{\mathcal{R}}_{jk}(u)
   \exp{ \{ - h(u) (2n_{j \uparrow}-1)(2n_{j \downarrow}-1) \}}.
\label{hubbardL}
\end{align}

Comparing eq.\ (\ref{hcom}) with the Yang-Baxter relation 
(\ref{YBRhubbard}),
we get a relation
\begin{align}
\nonumber\\
&\mathcal{I}_1(u) \mathcal{I}_2(v)
\mathcal{R}^{\rm h}_{12}(u,v) 
\mathcal{I}_1(u)^{-1} \mathcal{I}_2(v)^{-1} \nonumber \\
&\hspace{15mm}
= \alpha \bar{\mathcal{R}}_{12}(u-v) 
+ \beta (2n_{1 \uparrow}-1)(2n_{1 \downarrow}-1) \bar{\mathcal{R}}_{12}(u+v),
\label{rllcondition}
\end{align}
where
\begin{align}
&\hspace{5mm}
\mathcal{I}_j(u) 
= \exp\{ - h(u) (2n_{j \uparrow}-1)(2n_{j \downarrow}-1)\}.
\label{S}
\end{align}
From (\ref{rllcondition}), we obtain two conditions 
\begin{align}
  \dfrac{\beta a(u+v)}{\alpha a(u-v)}
       =  - \tanh (h(u)-h(v)) ,
%\nonumber\\
\hspace{5mm}
  \dfrac{\beta b(u+v)}{\alpha b(u-v)}
       = - \tanh (h(u)+h(v)).
\label{hrel}
\end{align}
which give the ratio of $\alpha$ to $\beta$
and constraints on $h(u)$ and ${h(v)}$.
The constraints are given more explicitly by
\begin{align}
\dfrac{\sinh 2h(u)}{\sin 2u} = \dfrac{\sinh 2h(v)}{\sin 2 v} = \frac{U}{4}.
\label{hcond}
\end{align}

To sum up, we have obtained the fermionic ${R}$-operator for the 
1D Hubbard model
\begin{align}
\mathcal{R}^{\rm h}_{12}(u,v)  =&
\mathcal{R}^{(\uparrow)}_{12}(u-v) 
\mathcal{R}^{(\downarrow)}_{12}(u-v)
-  \dfrac{\cos (u-v)}{\cos (u+v)} \tanh (h(u) - h(v)) \nonumber \\
& \times \mathcal{R}^{(\uparrow)}_{12}(u+v) 
\mathcal{R}^{(\downarrow)}_{12}(u+v)
(2n_{1 \uparrow}-1)(2n_{1 \downarrow}-1),
\label{fermionicRhubbard}
\end{align}
which satisfies the Yang-Baxter relation (\ref{YBRhubbard}). 
One of the remarkable properties of the fermionic ${R}$-operator 
(\ref{fermionicRhubbard}) is that its dependence on 
the the spectral parameters is not a difference type ${u-v}$. 
We call it the ``non-additive" property. 
The non-additive property of the fermionic ${R}$-operator 
(\ref{fermionicRhubbard}) allows us to generalize 
the Hamiltonian of the 1D Hubbard model (see \S 5).  

From (\ref{YBRhubbard}) we get the global Yang-Baxter relation,
\begin{align}
             \mathcal{R}^{\rm h}_{12}(u,v)       
     \mathcal{T}_1(u) \mathcal{T}_2(v)
           =&      \mathcal{T}_2(v) \mathcal{T}_1(u)
     \mathcal{R}^{\rm h}_{12}(u,v) 
\label{globalHubYBR}
\end{align}
for the monodromy operator
\begin{align}
     \mathcal{T}_a(u) =  \mathcal{L}_{aN}(u) \ldots \mathcal{L}_{a1}(u).
\label{monodromyhubbard}
\end{align}
Now we define the supertrace $\fstr^{\rm h}$ for the 1D Hubbard model as 
\begin{align}
     \fstr^{\rm h}_a X = {}_a \langle  0| X | 0 \rangle_a
           - {}_a \langle  \uparrow | X | \uparrow \rangle_a
           - {}_a \langle  \downarrow | X | \downarrow \rangle_a
           + {}_a \langle  \downarrow \uparrow | X | \uparrow \downarrow \rangle_a,
\end{align}
where the auxiliary fermion Fock space is given by
\begin{align}
c_{a \sigma} | 0 \rangle_a = 0, \hspace{5mm} | \uparrow \rangle_a &=  c_{a \uparrow}^{\dagger} | 0 \rangle_a, \hspace{5mm}
| \downarrow \rangle_a =  c_{a \downarrow}^{\dagger} | 0 \rangle_a, \hspace{5mm}| \uparrow \downarrow \rangle_a =  c_{a \uparrow}^{\dagger} c_{a \downarrow}^{\dagger} | 0 \rangle_a, \nonumber \\
{}_{a} \langle 0 | = (| 0 \rangle_a)^{\dagger}, \hspace{5mm} 
{}_{a} \langle \uparrow | &= {}_{a} \langle 0 | c_{a\uparrow}, \hspace{5mm}
{}_{a} \langle \downarrow | = {}_{a} \langle 0 | c_{a\downarrow}, \hspace{5mm}
{}_{a} \langle \downarrow \uparrow| = {}_{a} \langle 0 | c_{a\downarrow} c_{a\uparrow}.
\end{align}

Then the transfer operator of the 1D Hubbard model is defined by
\begin{align}
\tau(u) = \fstr^{\rm h}_a \mathcal{T}_{a}(u).
\end{align}
From the global Yang-Baxter relation (\ref{globalHubYBR}), we can show the existence of a commuting family of the transfer operators 
\begin{align}
    [ \tau(u) , \tau(v)] = 0,
\end{align}
which proves the integrability of the model.

Using the relations ${h(0) = 0}$ and ${h'(0) = U/4}$,
we can obtain the Hamiltonian of the 1D Hubbard model 
as a logarithmic derivative of the transfer operator,
\begin{align}
\mathcal{H} &= - \tau(0)^{-1} 
              \dfrac{{\rm d}}{{\rm d} u} \tau(u) \Big|_{u=0} 
\nonumber\\
            &=  - \sum_{j=1}^N 
               \sum_{\sigma = \uparrow \downarrow} 
                  \mathcal{H}^{{\rm XX}(\sigma)}_{j,j+1}  
        + \dfrac{U}{4} \sum_{j=1}^{N} 
        (2n_{j \uparrow} - 1)(2n_{j \downarrow} -1),
\end{align}
where the PBC for the fermion operators (\ref{pbcfermion}) is satisfied.

In this way we have proved the integrability of 
the 1D Hubbard model using the fermionic ${R}$-operator. 
It should be emphasized that we have not used the Jordan-Wigner 
transformation at all, which changes the boundary condition 
and the symmetry of the model. \cite{shiroishi4} 
 We shall see in \S 6 that the fermionic ${R}$-operator 
(\ref{fermionicRhubbard}) naturally enjoys the ${SO(4)}$ symmetry.

%%%%%%%%%%%%%%%%%%%%%%%%%%%%%%%%%%%%%%%%%%%%%%%%%%%%%%%%%%%%%%%%%%%

\section{Yang-Baxter Equation for the New Fermionic $R$-Operator}
\setcounter{equation}{0}
\renewcommand{\theequation}{5.\arabic{equation}}
In the previous section we have shown that 
the fermionic ${R}$-operator (\ref{fermionicRhubbard}) 
satisfies the Yang-Baxter relation (\ref{YBRhubbard}) 
with the ${L}$-operator (\ref{hubbardL}).  
It is natural to expect that the fermionic ${R}$-operator itself 
fulfills the Yang-Baxter equation (YBE),
\begin{align}
    \mathcal{R}_{12}^{\rm h}(u_1,u_2)
  \mathcal{R}_{13}^{\rm h}(u_1,u_3) \mathcal{R}_{23}^{\rm h} (u_2,u_3)
=   \mathcal{R}_{23}^{\rm h} (u_2,u_3)\mathcal{R}_{13}^{\rm h} (u_1,u_3)
  \mathcal{R}_{12}^{\rm h} (u_1,u_2).
\label{syang}
\end{align}

In the case of the coupled spin model, the YBE of 
the ${R}$-matrix was proved in ref.\ \cite{shiroishi1} 
using the tetrahedral Zamolodchikov algebra (TZA). 
\cite{korepanov,shiroishi3} 
In this section, we introduce the TZA for the fermionic ${R}$-operator 
and proves the YBE (\ref{syang}).

The TZA is defined by the following set of relations 
among operators
${\mathcal{L}_{jk}^{0}}$ and ${\mathcal{L}_{jk}^{1}}$,
\begin{align}
\mathcal{L}_{12}^{a} \mathcal{L}_{13}^{b} 
\mathcal{L}_{23}^{c} &= \sum_{d,e,f=0,1} S_{def}^{abc} \mathcal{L}_{23}^{a} \mathcal{L}_{13}^{b} 
\mathcal{L}_{12}^{c}, \hspace{2cm} a, \ldots, f =0,1,
\label{TZA}
\end{align}
where ${S_{def}^{abc}}$ are some scalar coefficients.

Let us take ${\mathcal{L}_{jk}^{(0)}}$ and ${\mathcal{L}_{jk}^{(1)}}$ as
\begin{align}
\mathcal{L}_{jk}^{0 (\sigma)} &= \mathcal{R}_{jk}^{(\sigma)}(u_j-u_k), \nonumber \\
\mathcal{L}_{jk}^{1 (\sigma)} &= \mathcal{R}_{jk}^{(\sigma)}(u_j+u_k) (2 n_{j\sigma} - 1), \hspace{2cm} \sigma = \uparrow \downarrow, \label{L0L1}
\end{align}
where ${\mathcal{R}_{jk}^{(\uparrow)}(u)}$ and ${\mathcal{R}_{jk}^{(\downarrow)}(u)}$ are the fermionic ${R}$-operators for the ${XX}$ fermion model as before.
Then we have found the following relations which give the TZA (\ref{TZA}),
\begin{align}
\mathcal{L}^{0 (\sigma)}_{12} \mathcal{L}^{0 (\sigma)}_{13} 
\mathcal{L}^{0 (\sigma)}_{23}
&= \mathcal{L}^{0 (\sigma)}_{23} \mathcal{L}^{0 (\sigma)}_{13} 
\mathcal{L}^{0 (\sigma)}_{12},
\hspace{5mm}
\mathcal{L}^{0 (\sigma)}_{12} \mathcal{L}^{1 (\sigma)}_{13} 
\mathcal{L}^{1 (\sigma)}_{23}
= \mathcal{L}^{1 (\sigma)}_{23} \mathcal{L}^{1 (\sigma)}_{13} 
\mathcal{L}^{0 (\sigma)}_{12}
\label{strd}
\\
\mathcal{L}^{1 (\sigma)}_{12} \mathcal{L}^{1 (\sigma)}_{13} 
\mathcal{L}^{0 (\sigma)}_{23}
&= \mathcal{L}^{0 (\sigma)}_{23} \mathcal{L}^{1 (\sigma)}_{13} 
\mathcal{L}^{1 (\sigma)}_{12},
\hspace{5mm}
\mathcal{L}^{1 (\sigma)}_{12} \mathcal{L}^{0 (\sigma)}_{13} 
\mathcal{L}^{1 (\sigma)}_{23}
= \mathcal{L}^{1 (\sigma)}_{23} \mathcal{L}^{0 (\sigma)}_{13} 
\mathcal{L}^{1 (\sigma)}_{12},
\label{dstrd}
\\
\mathcal{L}^{1 (\sigma)}_{12} \mathcal{L}^{1 (\sigma)}_{13} 
\mathcal{L}^{1 (\sigma)}_{23}
&=S_{001}^{111} 
%\dfrac{d_{12} c_{13}}{a_{12} b_{13}} 
\mathcal{L}^{1 (\sigma)}_{23} \mathcal{L}^{0 (\sigma)}_{13} 
\mathcal{L}^{0 (\sigma)}_{12}
+S_{010}^{111} 
%-\dfrac{d_{12} d_{23}}{a_{12} a_{23}}
\mathcal{L}^{0 (\sigma)}_{23} \mathcal{L}^{1 (\sigma)}_{13} 
\mathcal{L}^{0 (\sigma)}_{12}
+S_{100}^{111} 
%-\dfrac{d_{23} c_{13}}{a_{23} b_{13}}
\mathcal{L}^{0 (\sigma)}_{23} \mathcal{L}^{0 (\sigma)}_{13} 
\mathcal{L}^{1 (\sigma)}_{12},
%\nonumber
\\
\mathcal{L}^{0 (\sigma)}_{12} \mathcal{L}^{0 (\sigma)}_{13} 
\mathcal{L}^{1 (\sigma)}_{23}
&=S_{111}^{001}
%\dfrac{b_{12} a_{13}}{a'_{12} b'_{13}}
\mathcal{L}^{1 (\sigma)}_{23} \mathcal{L}^{1 (\sigma)}_{13} 
\mathcal{L}^{1 (\sigma)}_{12}
+S_{100}^{001}
%\dfrac{b_{12} b'_{23}}{a'_{12} a_{23}}
\mathcal{L}^{0 (\sigma)}_{23} \mathcal{L}^{0 (\sigma)}_{13} 
\mathcal{L}^{1 (\sigma)}_{12}
+S_{010}^{001}
%\dfrac{b'_{23} a_{13}}{a_{23} b'_{13}}
\mathcal{L}^{0 (\sigma)}_{23} \mathcal{L}^{1 (\sigma)}_{13} 
\mathcal{L}^{0 (\sigma)}_{12},
%\nonumber
\\
\mathcal{L}^{0 (\sigma)}_{12} \mathcal{L}^{1 (\sigma)}_{13} 
\mathcal{L}^{0 (\sigma)}_{23}
&=S_{111}^{010}
%\dfrac{b_{12} b_{23}}{a'_{12} a'_{23}}
\mathcal{L}^{1 (\sigma)}_{23} \mathcal{L}^{1 (\sigma)}_{13} 
\mathcal{L}^{1 (\sigma)}_{12}
+S_{100}^{010}
%\dfrac{b_{12} a'_{13}}{a'_{12} b_{13}}
\mathcal{L}^{0 (\sigma)}_{23} \mathcal{L}^{0 (\sigma)}_{13} 
\mathcal{L}^{1 (\sigma)}_{12}
+S_{001}^{010}
%\dfrac{b_{23} a'_{13}}{a'_{23} b_{13}}
\mathcal{L}^{1 (\sigma)}_{23} \mathcal{L}^{0 (\sigma)}_{13} 
\mathcal{L}^{0 (\sigma)}_{12},
%\nonumber
\\
\mathcal{L}^{1 (\sigma)}_{12} \mathcal{L}^{0 (\sigma)}_{13} 
\mathcal{L}^{0 (\sigma)}_{23}
&=S_{111}^{100}
%-\dfrac{b_{23} a_{13}}{a'_{23} b'_{13}}
\mathcal{L}^{1 (\sigma)}_{23} \mathcal{L}^{1 (\sigma)}_{13} 
\mathcal{L}^{1 (\sigma)}_{12}
+S_{010}^{100}
%\dfrac{b'_{12} a_{13}}{a_{12} b'_{13}}
\mathcal{L}^{0 (\sigma)}_{23} \mathcal{L}^{1 (\sigma)}_{13} 
\mathcal{L}^{0 (\sigma)}_{12}
+S_{001}^{100}
%-\dfrac{b'_{12} b_{23}}{a_{12} a'_{23}}
\mathcal{L}^{1 (\sigma)}_{23} \mathcal{L}^{0 (\sigma)}_{13} 
\mathcal{L}^{0 (\sigma)}_{12},
\label{lidel}
\end{align}
where ${\sigma = \uparrow}$ or ${\downarrow}$, and the coefficients
${S_{def}^{abc}}$ are given by
\begin{align}
S_{001}^{111} &= %\dfrac{d_{12}c_{13}}{a_{12}b_{13}}
\dfrac{\sin(u_1+u_2) \cos(u_1+u_3)}{\cos(u_1-u_2) \sin(u_1-u_3)},
\hspace{5mm}
S_{010}^{111} = %- \dfrac{d_{12}d_{23}}{a_{12}a_{23}}
- \dfrac{\sin(u_1+u_2) \sin(u_2+u_3)}{\cos(u_1-u_2) \cos(u_2-u_3)},
\nonumber\\
S_{100}^{111} &= % - \dfrac{d_{23}c_{13}}{a_{23}b_{13}}
- \dfrac{\sin(u_2+u_3) \cos(u_1+u_3)}{\cos(u_2-u_3) \sin(u_1-u_3)},
\hspace{5mm}
S_{111}^{001} = %\dfrac{b_{12}a_{13}}{c_{12}d_{13}}
\dfrac{\sin(u_1-u_2) \cos(u_1-u_3)}{\cos(u_1+u_2) \sin(u_1+u_3)},
\nonumber\\
S_{100}^{001} &= %\dfrac{b_{12}d_{23}}{c_{12}a_{23}}
\dfrac{\sin(u_1-u_2) \sin(u_2+u_3)}{\cos(u_1+u_2) \cos(u_2-u_3)},
\hspace{5mm}
S_{010}^{001} = %\dfrac{d_{23}a_{13}}{a_{23}d_{13}}
\dfrac{\sin(u_2+u_3) \cos(u_1-u_3)}{\cos(u_2-u_3) \sin(u_1+u_3)},
\nonumber\\
S_{111}^{010} &=
\dfrac{\sin(u_1-u_2) \sin(u_2-u_3)}{\cos(u_1+u_2) \cos(u_2+u_3)},
\hspace{5mm}
S_{100}^{010} =
\dfrac{\sin(u_1-u_2) \cos(u_1+u_3)}{\cos(u_1+u_2) \sin(u_1u_3)},
\nonumber\\
S_{001}^{010} &=
\dfrac{\sin(u_2-u_3) \cos(u_1+u_3)}{\cos(u_2+u_3) \sin(u_1-u_3)},
\hspace{5mm}
S_{111}^{100} =
- \dfrac{\sin(u_2-u_3) \cos(u_1-u_3)}{\cos(u_2+u_3) \sin(u_1+u_3)},
\nonumber\\
S_{010}^{100} &=
\dfrac{\sin(u_1+u_2) \cos(u_1-u_3)}{\cos(u_1-u_2) \sin(u_1+u_3)},
\hspace{5mm}
S_{001}^{100} =
- \dfrac{\sin(u_1+u_2) \sin(u_2-u_3)}{\cos(u_1-u_2) \cos(u_2+u_3)}.
\end{align}
Note that eq.\ (\ref{strd}) and eq.\ (\ref{dstrd}) are equivalent to the YBE (\ref{hstr}) and the DYBE (\ref{hdstr}) respectively. In this sense, the TZA (\ref{TZA}) can be regarded as a generalization of the YBE and the DYBE. 

We also remark that the products $\mathcal{L}^{a (\sigma)}_{12} \mathcal{L}^{b (\sigma)}_{13} \mathcal{L}^{c (\sigma)}_{23}$
${(a,b,c= 0,1)}$ are not linearly independent and satisfy the following relations, 
\begin{align}
\mathcal{L}^{0 (\sigma)}_{12} \mathcal{L}^{0 (\sigma)}_{13} \mathcal{L}^{0 (\sigma)}_{23}
&=x %-\dfrac{b_{13} a_{23}}{b'_{13} a'_{23}}
\mathcal{L}^{0 (\sigma)}_{12} \mathcal{L}^{1 (\sigma)}_{13} \mathcal{L}^{1 (\sigma)}_{23}
+y %\dfrac{a_{12} a_{23}}{a'_{12} a'_{23}}
\mathcal{L}^{1 (\sigma)}_{12} \mathcal{L}^{0 (\sigma)}_{13} \mathcal{L}^{1 (\sigma)}_{23}
+z %\dfrac{a_{12} a_{13}}{a'_{12} a'_{13}}
\mathcal{L}^{1 (\sigma)}_{12} \mathcal{L}^{1 (\sigma)}_{13} \mathcal{L}^{0 (\sigma)}_{23},
%\nonumber
 \label{linear1} \\
\mathcal{L}^{1 (\sigma)}_{12} \mathcal{L}^{1 (\sigma)}_{13} \mathcal{L}^{1 (\sigma)}_{23}
&=x'%-\dfrac{b'_{13} a'_{23}}{b_{13} a_{23}}
\mathcal{L}^{1 (\sigma)}_{12} \mathcal{L}^{0 (\sigma)}_{13} \mathcal{L}^{0 (\sigma)}_{23}
+y'%\dfrac{a'_{12} a'_{23}}{a_{12} a_{23}}
\mathcal{L}^{0 (\sigma)}_{12} \mathcal{L}^{1 (\sigma)}_{13} \mathcal{L}^{0 (\sigma)}_{23}
+z'%\dfrac{b'_{13} a'_{12}}{b_{13} a_{12}}
\mathcal{L}^{0 (\sigma)}_{12} \mathcal{L}^{0 (\sigma)}_{13}
\mathcal{L}^{1 (\sigma)}_{23}, 
\label{linear2}
\end{align}
where
\begin{align}
x &=
- \dfrac{\sin(u_1-u_3) \cos(u_2-u_3)}{\sin(u_1+u_3) \cos(u_2+u_3)},
\hspace{5mm}
y =
\dfrac{\cos(u_1-u_2) \cos(u_2-u_3)}{\cos(u_1+u_2) \cos(u_2+u_3)},
\nonumber\\
z &=
\dfrac{\cos(u_1-u_2) \cos(u_1-u_3)}{\cos(u_1+u_2) \cos(u_1+u_3)},
\hspace{5mm}
x' =
- \dfrac{\sin(u_1+u_3) \cos(u_2+u_3)}{\sin(u_1-u_3) \cos(u_2-u_3)},
\nonumber\\
y' &=
\dfrac{\cos(u_1+u_2) \cos(u_2+u_3)}{\cos(u_1-u_2) \cos(u_2-u_3)},
\hspace{5mm}
z' =
\dfrac{\sin(u_1+u_3) \cos(u_1+u_2)}{\sin(u_1-u_3) \cos(u_1-u_2)}.
\label{linearcoe}
\end{align}
These relations (\ref{strd}) - (\ref{linearcoe}) have been verified 
using mathematica.

Now taking into account the form of the fermionic ${R}$-operator 
(\ref{fermionicRhubbard}), 
we look for a solution of the YBE (\ref{syang}) in the form
\begin{align}
\mathcal{R}_{jk}^{\rm h} (u_j,u_k)
 = \mathcal{L}^{0 (\uparrow)}_{jk}(u_j,u_k) 
   \mathcal{L}^{0 (\downarrow)}_{jk}(u_j,u_k)
 + \alpha_{jk} \mathcal{L}^{1 (\uparrow)}_{jk}(u_j,u_k) 
   \mathcal{L}^{1 (\downarrow)}_{jk}(u_j,u_k).
\label{S1}
\end{align}
We substitute eq.\ (\ref{S1}) into the YBE (\ref{syang}). 
Using the TZA (\ref{TZA}) and the relations (\ref{linear1}) and (\ref{linear2}), we obtain the conditions 
\begin{align}
&\alpha_{13} \sin 2(u_1+u_3)
%\nonumber\\
=\alpha_{12} \sin 2(u_1+u_2) + \alpha_{23} \sin 2(u_2+u_3)
\nonumber\\
&=\dfrac{1}{\alpha_{12}} \sin 2(u_1-u_2) 
- \dfrac{1}{\alpha_{13}} \sin 2(u_1-u_3).
\label{scond}
\end{align} 
If we assume ${\alpha_{ij}}$ in the form
\begin{align}
\alpha_{jk}= - \dfrac{\cos(u_j-u_k)}{\cos(u_j+u_k)}
                   \tanh(h(u_j)-h(u_k)).
\end{align}
and impose the constraints
\begin{align}
\dfrac{\sinh 2 h(u_j)}{\sin 2 u_{j}} = \frac{U}{4}, \hspace{2cm} j=1,2,3,
\label{constraints}
\end{align}
then one can easily find that the conditions (\ref{scond}) are satisfied. 
In conclusion, the fermionic ${R}$-operator for the 1D Hubbard model,
\begin{align}
\mathcal{R}^{\rm h}_{jk}(u_j,u_k)  =&
\mathcal{R}^{(\uparrow)}_{jk}(u_j - u_k) 
\mathcal{R}^{(\downarrow)}_{jk}(u_j-u_k)
-  \dfrac{\cos (u_j-u_k)}{\cos (u_j+u_k)} \tanh (h(u_j) - h(u_k)) \nonumber \\
& \times \mathcal{R}^{(\uparrow)}_{jk}(u_j+u_k) 
\mathcal{R}^{(\downarrow)}_{jk}(u_j+u_k)
(2n_{j \uparrow}-1)(2n_{j \downarrow}-1),
\label{Rhubbard}
\end{align}
is a solution of the YBE (\ref{syang}) under the constraints 
(\ref{constraints}).

Besides the YBE (\ref{syang}), the fermionic ${R}$-operator 
(\ref{Rhubbard}) has the following properties
\begin{align}
&\mathcal{R}_{jk}^{\rm h}(u,0) = \dfrac{1}{\cosh h(u)} \mathcal{L}_{jk}(u), \label{fundamental} \\
&\mathcal{R}_{jk}^{\rm h}(u_0,u_0) = \mathcal{P}_{jk}, \label{permutator} \\
&\mathcal{R}_{jk}^{\rm h}(u_j,u_k) \mathcal{R}_{kj}^{\rm h}(u_k,u_j)
= \rho(u_j,u_k) {\bf 1},
\end{align}
where 
\begin{align}
\rho(u_j,u_k) = \cos^2 (u_j-u_k) \left\{ \cos^2 (u_j-u_k) - 
            \tanh^2 (h(u_j) -h(u_k)) \cos^2 (u_j + u_k) \right\}.
\end{align}
Here the permutation operator is defined by
\begin{align}
\mathcal{P}_{jk} &= \mathcal{P}_{jk}^{(\uparrow)} \mathcal{P}_{jk}^{(\downarrow)} \nonumber \\
\mathcal{P}_{jk}^{(\sigma)} &= 1 - (c^{\dagger}_{j \sigma} - c^{\dagger}_{k \sigma})(c_{j \sigma} - c_{k \sigma}), \hspace{2cm} \sigma = \uparrow \downarrow.
\end{align}
Due to eq.\ (\ref{fundamental}), we can recover the Yang-Baxter relation (\ref{YBRhubbard}) by putting ${u_3=0}$ in the YBE (\ref{syang}). 

Using the ${R}$-operator (\ref{Rhubbard}), 
we can introduce an inhomogeneous model as
\begin{align}
\mathcal{T}_{a}(u,\{u_j\}) = \mathcal{R}_{aN}^{\rm h} (u,u_N) \ldots \mathcal{R}^{\rm h}_{aj}(u,u_j) \ldots \mathcal{R}_{a1}^{\rm h} (u,u_1), 
\end{align}
where ${u_{j} \  (j=1, \ldots,N)}$ are the inhomogeneous parameters 
obeying the constraints
\begin{align}
\dfrac{\sinh 2 h(u_j)}{\sin 2 u_{j}} = \frac{U}{4}, \hspace{2cm} j=1,\ldots,N.
\end{align}
Then from the YBE (\ref{syang}), we have the global Yang-Baxter relation
\begin{align} 
    \mathcal{R}_{12}^{\rm h} (u,v)
  \mathcal{T}_{1}(u,\{u_j\}) \mathcal{T}_{2}(v,\{u_j\})
=&   \mathcal{T}_{2}(v,\{u_j\})\mathcal{T}_{1}(u,\{u_j\})
  \mathcal{R}_{12}^{\rm h} (u,v),
\label{globalYBE}
\end{align}
which leads to the commutativity
\begin{align}
\left[ \tau(u,\{u_j\}), \tau(v,\{u_j\}) \right] = 0.
\end{align}

Because of the non-additive property of 
the ${R}$-operator (\ref{Rhubbard}), the simplest choice 
\begin{align}
\mathcal{T}_{a}(u,u_{0}) = \mathcal{R}_{aN}^{\rm h} (u,u_0) \ldots 
\mathcal{R}^{\rm h}_{aj}(u,u_0) \ldots
\mathcal{R}^{\rm h}_{a1}(u,u_0)
\label{genetransfer}
\end{align}
still generalizes eq.\ (\ref{monodromyhubbard}). 
Actually we can derive a fermionic integrable Hamiltonian 
which generalize the 1D Hubbard model. 
Note that ${\tau(u_0,u_0)}$ is the shift operator 
regardless of the parameter ${u_0}$,
\begin{align}
\tau(u_0,u_0) &= \hat{U} \nonumber \\
              &\equiv \mathcal{P}_{12} \mathcal{P}_{23} 
                          \ldots \mathcal{P}_{N-1,N}. 
\end{align}
Then the logarithmic derivative of the generalized transfer operator yields a new fermionic Hamiltonian as follows,
\begin{align} 
\mathcal{H} =& - \tau(u_0,u_0)^{-1} \dfrac{\rm d}{{\rm d} u} \tau(u,u_0) \Big|_{u=u_0}  \nonumber \\
=& - \sum_{j=1}^N \sum_{\sigma=\uparrow \downarrow} 
  \left( c_{j \sigma}^{\dagger} c_{j+1 \sigma} 
+ c_{j+1 \sigma}^{\dagger} c_{j \sigma} \right) \nonumber \\
&+ \dfrac{U}{4 \cosh 2h(u_0)} \sum_{j=1}^{N}
  \Big\{
  \cos ^2 u_0 (2n_{j \uparrow}-1) - \sin ^2 u_0 (2n_{j+1 \uparrow}-1) \nonumber \\
 & \hspace{7cm} +\sin 2u_0
     (  c_{j+1 \uparrow}^{\dagger} c_{j \uparrow}
   -  c_{j \uparrow}^{\dagger}  c_{j+1 \uparrow}  )     
  \Big\}
\nonumber\\ 
 & \hspace{3cm} \times \Big\{ 
 \cos ^2 u_0 (2n_{j \downarrow}-1) - \sin ^2 u_0 (2n_{j+1 \downarrow}-1) \nonumber \\ 
 & \hspace{7cm} +\sin 2u_0
     (  c_{j+1 \downarrow}^{\dagger}  c_{j \downarrow}
   -  c_{j \downarrow}^{\dagger}  c_{j+1 \downarrow}  )   
  \Big\}.
\label{modifiedhub}
\end{align}
If we take ${u_0 = 0}$ or ${\pi/2}$, the Hamiltonian (\ref{modifiedhub}) reduces to 
the 1D Hubbard model (\ref{hamiltonian.hubbard}).
The local higher conserved operators can also be obtained by the formula
\begin{align}
I_n = \frac{{\rm d}^n}{{\rm d} u^n} \ln 
\left( \tau(u_0,u_0)^{-1} \tau(u,u_0)  \right) \Big|_{u=u_0}, \hspace{2cm}
n=1,2, \ldots,
\label{conserved}
\end{align}
where ${I_{1} = - \mathcal{H}}$. The meaning of the insertion of 
the term 
${\tau(u_0,u_0)^{-1}}$ in (\ref{conserved}) will be clarified when we discuss the ${SO(4)}$ symmetry in \S 6.

Remark that in terms of 
${\mathcal{\check{R}}_{jk}(u,u_0) \equiv \mathcal{P}_{jk} \mathcal{R}_{jk}(u,u_0)}$ we can express the Hamiltonian (\ref{modifiedhub}) as
\begin{align}
\mathcal{H} = - \sum_{j=1}^{N} \dfrac{\rm d}{{\rm d} u} \mathcal{\check{R}}_{jk}(u,u_0) \Big|_{u=u_0}.
\label{hamiltonianRcheck}
\end{align}
The higher conserved operators ${I_n}$ are also expressed 
in terms of ${\mathcal{\check{R}}_{jk}(u,u_0)}$, 
though it is very difficult to get their explicit forms.

%%%%%%%%%%%%%%%%%%%%%%%%%%%%%%%%%%%%%%%%%%%%%

\section{SO(4) Symmetry}
\setcounter{equation}{0}
\renewcommand{\theequation}{6.\arabic{equation}}
The ${SO(4) (= [ SU(2) \times SU(2)]/Z_{2})}$ symmetry is one of the most 
important properties of the Hubbard Hamiltonian 
(\ref{hamiltonian.hubbard}) \cite{heilmann,yang1,yang2,pernici,affleck}.
The two ${SU(2)}$ come from the spin ${SU(2)}$ 
\begin{align} 
S^{+} &= \sum_{j=1}^{N} c_{j\uparrow}^{\dagger} c_{j\downarrow}, \hspace{5mm} 
S^{-}  = \sum_{j=1}^{N} c_{j\downarrow}^{\dagger} c_{j\uparrow}, \hspace{5mm}
S^{z} = \frac{1}{2} \sum_{j=1}^{N} (n_{j\uparrow} - n_{j\downarrow})
\label{spinSU(2)}
\end{align}
and the charge ${SU(2)}$ (${\eta}$-pairing ${SU(2)}$) 
\begin{align} 
\eta^{+} &= \sum_{j=1}^{N} (-1)^{j} c_{j\uparrow}^{\dagger} 
c_{j\downarrow}^{\dagger}, \hspace{5mm}
\eta^{-}  = \sum_{j=1}^{N} (-1)^{j} c_{j\downarrow} c_{j\uparrow},
\hspace{5mm} 
\eta^{z} = \frac{1}{2} \sum_{j=1}^{N} (n_{j\uparrow}+n_{j\downarrow}-1).   
\label{eq.chargeSU(2)}
\end{align} 
We assume that the lattice has an even number of sites
under the PBC (\ref{pbcfermion}). Then the six generators above 
commute with the Hamiltonian 
(\ref{hamiltonian.hubbard}).

In ref. \cite{murakami,shiroishi4}, 
the ${SO(4)}$ symmetry of the fermionic
transfer matrix was discussed based on the fermionic formulation by 
Olmedilla {\it et al.}. \cite{wadati2} In the following we discuss 
the ${SO(4)}$ symmetry of the fermionic ${R}$-operator and the generalized 
transfer operator (\ref{genetransfer}) 
using the similar discussion (c.f. ref. \cite{takhtajan}).  
For this purpose, 
we introduce the local generators of the ${SO(4)}$ algebra as follows,
\begin{align}
S_{j}^{+} &= c_{j\uparrow}^{\dagger} c_{j\downarrow}, \hspace{5mm} 
S_{j}^{-} = c_{j\downarrow}^{\dagger} c_{j\uparrow}, \hspace{5mm} 
S_{j}^{z} = \frac{1}{2}(n_{j\uparrow} - n_{j\downarrow}), \nonumber \\
\eta_{j}^{+} &= c_{j\uparrow}^{\dagger} c_{j\downarrow}^{\dagger}, \hspace{5mm}
\eta_{j}^{-}  = c_{j\downarrow} c_{j\uparrow}, \hspace{5mm} 
\eta_{j}^{z}  = \frac{1}{2}(n_{j\uparrow}+n_{j\downarrow}-1). 
\end{align}
The ${SU(2)}$ symmetry of the fermionic ${R}$-operator for 
the 1D Hubbard model (\ref{Rhubbard}) is represented 
by the following commutation relations,
\begin{align}
&\Big[ \mathcal{R}_{jk}^{\rm h}(u,v), S_{j}^{+} + S_{k}^{+} \Big]=0, \hspace{5mm}
\Big[ \mathcal{R}_{jk}^{\rm h}(u,v), S_{j}^{-} + S_{k}^{-} \Big]=0, \nonumber \\
&\Big[ \mathcal{R}_{jk}^{\rm h}(u,v), S_{j}^{z} + S_{k}^{z} \Big]=0.
\label{spinR}
\end{align}
It is easy to see that the relations (\ref{spinR}) extend to the ${SU(2)}$
symmetry of the monodromy operator 
\begin{align} 
\mathcal{T}_{a}(u,u_0) = \mathcal{R}_{aN}^{\rm h}(u,u_0) 
\ldots \mathcal{R}_{a1}^{\rm h}(u,u_0) \nonumber 
\end{align}
as
\begin{align}
\Big[ \mathcal{T}_{a}(u,u_0), S_{a}^{\alpha} + S^{\alpha} \Big]=0,
\hspace{3cm} \alpha= \pm, z.
\label{spinmonodromy}
\end{align}
Taking the supertrace ${{\rm Str}_{a}}$ of (\ref{spinmonodromy}), we obtain 
\begin{align}
\Big[ \tau(u,u_0), S^{\alpha} \Big]=0,
\hspace{4cm} \alpha= \pm, z,
\label{spintransfer}
\end{align}
which shows that the transfer operator ${\tau(u,u_0)}$ is the spin ${SU(2)}$
invariant.

Next we consider the charge ${SU(2)}$ symmetry. For the generator 
${\eta_{j}^{z}}$, the same relation as (\ref{spinR}) holds,
\begin{align}
\Big[ \mathcal{R}_{jk}^{\rm h}(u,v), \eta_{j}^{z} + \eta_{k}^{z} \Big]=0,
\end{align}
which extends to the symmetry of the monodromy operator and the transfer
operator as, 
\begin{align}
\Big[ \mathcal{T}_{a}(u,u_0), \eta_{a}^{z} + \eta^{z} \Big]=0, 
\hspace{5mm}
\Big[ \tau(u,u_0), \eta^{z} \Big]=0.
\end{align}
As for the other generators ${\eta_{j}^{\pm}}$, we do not have the relations
like (\ref{spinR}). Instead, we have discovered the following anti-commuting 
relation for ${\eta_{j}^{\pm}}$,
\begin{align}
\Big\{ \mathcal{R}_{jk}^{\rm h}(u,v), \eta_{j}^{\pm} - \eta_{k}^{\pm} \Big\} = 0. 
\label{chargeR}
\end{align}

The local relation (\ref{chargeR}) extends to the identities for 
the monodromy operator,
\begin{align}
\Big[ \mathcal{T}_{a}(u,u_0), \eta_{a}^{\pm} \Big]
+ \Big\{ \mathcal{T}_{a}(u,u_0), \eta^{\pm} \Big\} = 0,
\label{chargepm}
\end{align}
which can be proved as follows,
\begin{align}   
&\mathcal{T}_{a}(u,u_0) \left( \eta_{a}^{\pm} + \eta^{\pm} \right)
\nonumber\\
&=\mathcal{R}_{aN}^{\rm h}(u,u_0) \ldots \mathcal{R}_{a2}(u,u_0)  
\mathcal{R}_{a1}^{\rm h}(u,u_0)  
(\eta_a^{\pm} - \eta_1^{\pm} + \eta_2^{\pm} \ldots + \eta_N^{\pm})
\nonumber\\
&= \mathcal{R}_{aN}^{\rm h}(u,u_0) \ldots \mathcal{R}_{a2}^{\rm h}(u,u_0)
(-\eta_a^{\pm} + \eta_1^{\pm} + \eta_2^{\pm} \ldots + \eta_N^{\pm})  
\mathcal{R}_{a1}^{\rm h}(u,u_0) 
\nonumber\\
&= \ldots 
\nonumber\\
&= (\eta_a^{\pm} + \eta_1^{\pm} - \eta_2^{\pm} \ldots - \eta_N^{\pm})  
\mathcal{R}_{aN}^{\rm h} (u,u_0) \ldots \mathcal{R}_{a2}^{\rm h} (u,u_0)  
\mathcal{R}_{a1}^{\rm h} (u,u_0) \nonumber \\
&= (\eta_{a}^{\pm} - \eta^{\pm}) \mathcal{T}_{a}(u,u_0).  
\end{align} 
Here we have used the fact that ${N}$ is even.

Taking the supertrace of (\ref{chargepm}), we obtain
\begin{align}
\Big\{ \tau(u,u_0), \eta^{\pm} \Big\} = 0.
\label{anticummute}
\end{align}
Namely the transfer operator anti-commutes with the generators ${\eta^{\pm}}$.
Thus contrary to the naive anticipation, the transfer operator itself does not
commute with the generators of the charge ${SU(2)}$ symmetry. 
However a combination ${ \tau(u_0,u_0)^{-1} \tau(u,u_0)}$ 
commutes with ${\eta^{\pm}}$
\begin{align}
\Big[   \tau(u_0,u_0)^{-1} \tau(u,u_0), \eta^{\pm} \Big] =0,
\end{align} 
due to the relation 
\begin{align}
\Big\{ \tau(u_0,u_0), \eta^{\pm} \Big\} = 0.
\label{shifteta}
\end{align} 

Of course, ${\tau(u_0,u_0)^{-1} \tau(u,u_0) }$ commutes with 
the other generators of the ${SO(4)}$ algebra. 
Since we have defined the local conserved operators by eq. (\ref{conserved}),
\begin{align}
I_n = \frac{{\rm d}^n}{{\rm d} u^n} \ln 
\left( \tau(u_0,u_0)^{-1} \tau(u,u_0)  \right) \Big|_{u=u_0}, \hspace{2cm}
n=1,2, \ldots
\end{align}
it follows that they are all ${SO(4)}$ invariant. 
In particular, we have proved that 
a generalized Hubbard model (\ref{modifiedhub}) has the 
${SO(4)}$ symmetry. It is also possible 
to prove the ${SO(4)}$ invariance of 
the Hamiltonian (\ref{modifiedhub}) directly 
using the expression (\ref{hamiltonianRcheck}) and the commutation relations 
\begin{align}
&\Big[ \mathcal{\check{R}}_{jk}^{\rm h}(u,v), S_{j}^{\alpha} + S_{k}^{\alpha} \Big]=0, \hspace{4cm} \alpha = \pm,z, \nonumber \\
&\Big[ \mathcal{\check{R}}_{jk}^{\rm h}(u,v), 
\eta_{j}^{z} + \eta_{k}^{z} \Big]
=0, \hspace{5mm}
\Big[ \mathcal{\check{R}}_{jk}^{\rm h}(u,v), \eta_{j}^{\pm} - \eta_{k}^{\pm} \Big]
=0.
\end{align} 
where ${\check{\mathcal{R}}_{jk}^{\rm h}(u,v) = \mathcal{P}_{jk} \mathcal{R}_{jk}^{\rm h}(u,v)}$ as before.

The relation (\ref{shifteta}) means that the shift
 operator ${\hat{U}=\tau(u_0,u_0)}$  anticummutes with ${\eta^{\pm}}$
\begin{align}
\hat{U} \eta^{\pm} = - \eta^{\pm} \hat{U}. 
\label{shift2}
\end{align}
If we define the lattice momentum operator ${\hat{\Pi}}$ 
\cite{takhtajan,murakami} by
\begin{align} 
\hat{U} = \exp {\rm i} \hat{\Pi},  
\label{momentum}
\end{align} 
the relation (\ref{shift2}) implies that
\begin{align}
\hat{\Pi} \eta^{\pm} - \eta^{\pm} \hat{\Pi} =  \pi \eta^{\pm}.
\label{pichange}
\end{align}
This indicates that the generators ${\eta^{\pm}}$ changes the momentum
eigenvalue by ${\pi}$. \cite{yang1,murakami} Note that the lattice
momentum operator ${\hat{\Pi}}$ is defined 
in (\ref{momentum}) mod ${2 \pi}$.  

The lattice momentum operator ${\hat{\Pi}}$ is also a conserved 
operator
\begin{align}
\left[ \hat{\Pi}, I_n \right] = 0, 
\end{align}
but it is not contained in ${\{I_{n}\}}$. Actually ${\hat{\Pi}}$
is a non-local operator (we refer ref. \cite{murakami} for 
the detailed discussions on ${\hat{\Pi}}$). Furthermore the 
relation (\ref{pichange}) suggests that it does not have 
the ${SO(4)}$ symmetry.

%%%%%%%%%%%%%%%%%%%%%%%%%%%%%%%%%%%%%%%%%%%%%%%%%%%%%%%%%%%%

\section{Concluding Remarks}
\setcounter{equation}{0}
\renewcommand{\theequation}{7.\arabic{equation}}
In this paper, we have presented a new fermionic approach 
for the integrability of the 1D Hubbard model. The central 
object in our approach is the fermionic ${R}$-operator, which
 we introduced recently to discuss the integrability of the ${XXZ}$
fermion model. We have shown that the fermionic ${R}$-
operator can be generalized to the ${XYZ}$ fermion model. 
It gives an operator solution of the Yang-Baxter equation (YBE). 
We also considered the decorated Yang-Baxter equation (DYBE) for
the fermionic ${R}$-operator and found that the free-fermion condition 
is necessary. In other words, the fermionic ${R}$-operator related 
to the ${XY}$ fermion model fulfills the YBE and the DYBE. 

The 1D Hubbard model can be regarded as two ${XX}$ fermion models 
with an interaction term.  Using the YBE and the DYBE 
for the ${XX}$ fermion model, we have established 
a new kind of the Yang-Baxter relation for the 1D Hubbard model. 
This gives an alternative proof for the integrability of the 1D Hubbard model. 
We have obtained the fermionic ${R}$-operator 
for the 1D Hubbard model and confirmed that it satisfies the YBE. 
The connection between the fermionic ${R}$-operator and the
graded Yang-Baxter relation for the 1D Hubbard model 
\cite{wadati2} will be discussed in a separate paper. \cite{shiroishi5}
 
From the fermionic ${R}$-operator of the 1D Hubbard model, we have 
constructed a new fermionic transfer operator. Because of the non-additive
property of the ${R}$-operator, the transfer operator generates 
a fermionic integrable Hamiltonian (\ref{modifiedhub})
which generalize the 1D Hubbard model.

One of the advantages of our fermionic approach is that we can 
discuss the ${SO(4)}$ symmetry at the level of the (local) fermionic 
${R}$-operator, which immediately leads to the ${SO(4)}$ symmetry of
all the local conserved operators derived from the transfer operator. 
It is known that the ${SO(4)}$ symmetry of the 1D Hubbard 
model is extended to the Yangian symmetry on the infinite lattice 
\cite{uglov,murakami2}. It is interesting to investigate 
the relation between the fermionic 
${R}$-operator and the Yangian symmetry. 

\vspace{5mm}
%%%%%%%%%%%%%%%%%%%%%%%%%%%%%%%%%%%%%%%%%%%%%%%%%%%%%%%%%%%%
\begin{center}
{\bf Acknowledgment}
\end{center}
The authors are grateful to Y. Komori
for useful discussions and careful reading of the manuscript. 
This work 
is in part supported by Grant-in-Aid for 
JSPJ Fellows from the
Ministry of Education, Science, Sports and Culture of Japan. 

\vspace{30pt}

\begin{flushleft}
APPENDIX A: A Simple Derivation of the Fermionic ${R}$-Operator 
\end{flushleft}
\setcounter{equation}{0}
\renewcommand{\theequation}{A.\arabic{equation}}
The ${R}$-matrix of the ${XYZ}$ spin models in terms of the Pauli
spin matrix ($\sigma^{+}_{j}, \sigma^{-}_{j}$ and 
$\sigma_{j}^{z}$) is given by
\begin{align}
R_{jk}(u) =& a(u) \dfrac{1 + \sigma^z_j \sigma^z_k}{2}
            + b(u) \dfrac{1 - \sigma^z_j \sigma^z_k}{2}
            + c(u) (\sigma^+_j \sigma^-_k + \sigma^+_k \sigma^-_j)
            \nonumber\\
           & + d(u) (\sigma^+_j \sigma^+_k + \sigma^-_j \sigma^-_k),  
\label{8V}
\end{align}
where the Boltzmann weights ${a(u), b(u), c(u)}$ and ${d(u)}$ are the same as
 (\ref{baxterpara}). The ${R}$-matrix (\ref{8V}) satisfies the Yang-Baxter equation
\begin{align}
R_{12}(u-v) R_{13}(u) R_{23}(v) = R_{23}(v) R_{13}(u) R_{12}(u-v).
\label{YBERmatrix}
\end{align}
We like to obtain the fermionic ${R}$-operator (\ref{R}) 
from the ${R}$-matrix (\ref{8V}) by means of 
the Jordan-Wigner transformation \cite{lieb2}
\begin{align}
\sigma_{j}^{-} &= \exp \left\{ {\rm i} \pi 
\sum_{k=1}^{j-1} n_{k} \right\} c_{j}, \nonumber \\
\sigma_{j}^{+} &= \exp \left\{ -{\rm i} \pi 
\sum_{k=1}^{j-1} n_{k} \right\} c_{j}^{\dagger},
 \label{JWT}
\end{align}
However it is not possible to apply the Jordan-Wigner transformation directly
to (\ref{8V}) because of the non-local term ${R_{13}(u)}$ in the Yang-Baxter equation (\ref{YBERmatrix}). For this reason, we consider an equivalent ${R}$-matrix ${\check{R}_{jk}(u)}$
\begin{align}
\check{R}_{jk}(u) &  \equiv P_{jk} R_{jk}(u)  \nonumber \\
= & a(u) \dfrac{1 + \sigma^z_j \sigma^z_k}{2}
           + c(u) \dfrac{1 - \sigma^z_j \sigma^z_k}{2}
           + b(u) (\sigma^+_j \sigma^-_k 
                     + \sigma^-_j \sigma^+_k)  
\nonumber\\
           & + d(u) (\sigma^+_j \sigma^+_k 
                      + \sigma^-_j \sigma^-_k),
\label{checkRmatrix}
\end{align}
which satisfies the Yang-Baxter equation in the form
\begin{align}
\check{R}_{12}(u-v) \check{R}_{23}(u) \check{R}_{12}(v) 
= \check{R}_{23}(v) \check{R}_{12}(u) \check{R}_{23}(u-v).
\label{YBEcheckRmatrix}
\end{align}
Here ${P_{jk}}$ is the permutation matrix
\begin{align}
P_{jk} = \dfrac{1 + \sigma^z_j \sigma^z_k}{2}
                  + \sigma^+_j \sigma^-_k 
                  + \sigma^-_j \sigma^+_k.  
\end{align}
        
Now we can apply the Jordan-Wigner transformation (\ref{JWT}) to
(\ref{checkRmatrix}) and (\ref{YBEcheckRmatrix}). Then we obtain
a fermionic solution
\begin{align}
       \check{\mathcal{R}}_{jk}(u) =& a(u) \{ n_j n_k + 
(1-n_j)(1-n_k)\}
           + c(u) \{n_j (1-n_k) + (1-n_j) n_k \}
\nonumber\\    
           +  & b(u) (c_j^{\dagger} c_k + c_k^{\dagger} c_j)
           + d(u) (c_j^{\dagger} c_k^{\dagger} - c_j c_k),
\label{fermioncheckR}
\end{align}
satisfying
\begin{align}
\check{\mathcal{R}}_{12}(u-v) \check{\mathcal{R}}_{23}(u) 
\check{\mathcal{R}}_{12}(v) 
= \check{\mathcal{R}}_{23}(v) \check{\mathcal{R}}_{12}(u) 
\check{\mathcal{R}}_{23}(u-v).
\label{fermionYBEcheckRmatrix}
\end{align}
In the derivation, the following transformation laws are 
of particular importance,
\begin{align}
&\sigma^{+}_{j} \sigma^{+}_{j+1} \longrightarrow c_{j}^{\dagger} c_{j+1}^{\dagger}, \nonumber \\
&\sigma^{-}_{j} \sigma^{-}_{j+1} \longrightarrow - c_{j} c_{j+1}.
\end{align} 
Strictly speaking, ${\check{\mathcal{R}}_{jk}(u)}$ is defined in 
(\ref{fermioncheckR}) only for ${k=j+1}$. However we extends its definition to arbitrary ${j}$ and ${k}$ by the formula 
(\ref{fermioncheckR}).

Finally multiplying the products of the fermionic permutation operators
(\ref{fermionicpermutation})
\begin{align}
\mathcal{P}_{12} \mathcal{P}_{13} \mathcal{P}_{23}
= \mathcal{P}_{23} \mathcal{P}_{13} \mathcal{P}_{12},
\label{permutatorYBE}
\end{align}
to (\ref{fermioncheckR}) from the left,
we obtain 
\begin{align}
\mathcal{R}_{12}(u-v) \mathcal{R}_{13}(u) 
\mathcal{R}_{23}(v) 
= \mathcal{R}_{23}(v) \mathcal{R}_{13}(u) 
\mathcal{R}_{12}(u-v),
\end{align}
with
\begin{align}
\mathcal{R}_{jk}(u) &= \mathcal{P}_{jk} \check{\mathcal{R}}_{jk}(u)  \nonumber \\
                    &= a(u) \{ -n_j n_{k} + (1-n_j)(1-n_{k})\}
           + b(u) \{n_j (1-n_k) + (1-n_j) n_k\}
\nonumber\\    
            & +c(u) (c_j^{\dagger} c_k + c_k^{\dagger} c_j)
           - d(u) (c_j^{\dagger} c_k^{\dagger} + c_j c_k).
\end{align}
This is identical to the fermionic ${R}$-operator (\ref{R}).
In this way we can obtain the fermionic ${R}$-operator (\ref{R}) from the ${R}$-matrix of the ${XYZ}$ spin model.

Just in the same way, the fermionic ${R}$-operator for the 1D
Hubbard model can be obtained from the ${R}$-matrix of the coupled spin model (\ref{coupledspin}).

\end{document}